\documentclass[preprint,nofootinbib,12pt]{revtex4}

\usepackage{graphicx}
\usepackage{psfrag}

\newcommand{\ket}[1]{\left|#1\right\rangle}

\newcommand{\vev}[1]{\left\langle #1 \right\rangle}

\newcommand{\lqcd}{\Lambda_{QCD}}
\newcommand{\GeV}{{\rm GeV}}
\newcommand{\MeV}{{\rm MeV}}
\newcommand{\TeV}{{\rm TeV}}
\newcommand{\eV}{{\rm eV}}
\newcommand{\ps}{{\rm ps}}

\newcommand{\beq}{\begin{equation}}
\newcommand{\eeq}{\end{equation}}
\newcommand{\beqa}{\begin{eqnarray}}
\newcommand{\eeqa}{\end{eqnarray}}

\newcommand{\lsim}{\mathrel{\rlap{\lower4pt\hbox{\hskip1pt$\sim$}}
    \raise1pt\hbox{$<$}}}         %less than or approx. symbol
\newcommand{\gsim}{\mathrel{\rlap{\lower4pt\hbox{\hskip1pt$\sim$}}
    \raise1pt\hbox{$>$}}}         %greater than or approx. symbol

\begin{document}

%\preprint{{\vbox{\hbox{}\hbox{}\hbox{}
%    \hbox{WIS/10/05-May-DPP}
%\hbox{hep-ph/yymmnnn}}}}

\vspace*{.0cm}

\title{Hadronization, spin, and lifetimes}

\author{Yuval Grossman}\email{yuvalg@lepp.cornell.edu}
\author{Itay Nachshon}\email{in36@lepp.cornell.edu}

\affiliation{\vspace*{4mm}Institute for High Energy Phenomenology\\
Newman Laboratory of Elementary Particle Physics\\
Cornell University, Ithaca, NY 14853, USA\vspace*{6mm}}

%\date{\today}
%\pacs{}

\vspace{1cm}
\begin{abstract}
Measurements of lifetimes can be done in two ways. For very short
lived particles, the width can be measured. For long lived ones, the
lifetime can be directly measured, for example, using a displaced
vertex. Practically, the lifetime cannot be extracted for particles
with intermediate lifetimes. We show that for such cases information
about the lifetime can be extracted for heavy colored particles that
can be produced with known polarization. For example, a $t$-like
particle with intermediate lifetime hadronizes into a superposition of
the lowest two hadronic states, $T^*$ and $T$ (the equivalent of $B^*$
and $B$). Depolarization effects are governed by time scales that are
much longer than the hadronization time scale, $\lqcd^{-1}$. After a
time of order $1/\Delta m$, with $\Delta m \equiv m(T^*)-m(T)$, half
of the initial polarization is lost. The polarization is totally lost
after a time of order $1/\Gamma_{\gamma}$, with $\Gamma_{\gamma}=
\Gamma(T^*\to T\gamma)$. Thus, by comparing the initial and final
polarization, we get information on the particle's lifetime.

\end{abstract}

\maketitle

%%%%%%%%%%%%%%%%%%%%%%%%%%%%
\section{Introduction}
There are strong motivations to hope that the LHC will find new
particles with electroweak scale masses. Once such a new particle is
discovered, the first task will be to determine its properties, in
particular, mass, charges, spin and lifetime. Clearly, such
determinations require much larger statistics than what is needed for
discoveries of new states. The hope is that eventually, with enough
data from the LHC and future machines, we will be able to determine
these properties for all the new particles.

In this work we concentrate on determining lifetimes. There are
basically two ways the lifetime of a particle can be determined. The
first is to directly measure its width. This method works when the
intrinsic width is larger than the experimental resolution. The
experimental sensitivity depends on many factors, like the mass of the
particle and its charge. Very roughly, for a particle with mass of a
few hundred $\GeV$, its width can be extracted when $\Gamma \gsim
1\;\GeV$. The other method to extract lifetimes is by looking for a
displaced secondary vertex. This can be done, very roughly, for $\tau
\gsim 1\;\ps$, that is, $\Gamma \lsim 10^{-4}
\;\eV$.  We see that there is a very large window,
\beq
10^9\;\eV \gsim \Gamma \gsim 10^{-4}\;\eV,
\eeq
which we denote as ``the problematic region,'' where lifetimes cannot
be extracted.

The fact that our ability to measure lifetimes is limited may not be a
problem. For example, in a generic SUSY model we expect the LSP to be
stable and all other super-particles to have widths that are larger
than $1\,\GeV$. This is the generic case in most available models of
physics beyond the SM; the unstable particles are very short lived
while other new particles are stable due to an exact symmetry.
There are, however, exceptions. There are well motivated models with new,
unstable particles with lifetimes that are much longer than the naive
ones, such that their widths are within the problematic region. This
is the case, for example, in $Z'$-mediated SUSY Breaking
\cite{Langacker:2008ip} (for the gluino and the
NLSP Wino), in split SUSY \cite{ArkaniHamed:2004fb} (for the gluino when
$m_s \lsim 1000\;\TeV$) and in GUTs in warped extra dimension 
\cite{Agashe:2004ci,Agashe:2004bm} (for the GUT partners).

Below we describe a way that, in principle, can tell us information
about a lifetime of a particle in the problematic region. The basic
idea is as follows. Consider a particle of mass $m$ that is not a
singlet of SU(3)$_C$ nor of the Lorentz group. Thus, if its lifetime
is longer than the QCD scale, it hadronizes before it decays.  If the
particle is produced polarized, the fact that it is hadronizes could
eventually reduce its initial polarization. In cases where the
polarization can be measured and compared to the expected one, we can
extract the amount of depolarization. Knowing the time scale
associated with the loss of polarization, make it possible to
determine if the lifetime is larger or smaller than that time scale.

We note that we expect the new physics not to conserve parity. For
example, in models with extra dimensions, the KK tower of the
right-handed and left handed quarks have different masses and
different couplings to the light fermions. Therefore, it is very
likely that these new particles will be produced with high degree of
polarization, and decay in a way that can be used to measure this
polarization.

The loss of polarization takes place at time scales much longer than
the hadronization scale \cite{Falk:1993rf}. The typical time for
hadronization is $\lqcd^{-1}$ while for depolarization it is $m
\lqcd^{-2}$. This is similar to the case of the
hydrogen atom. The energy scale associated with depolarization of the
heavy proton is that of the hyperfine splitting and it is suppressed
by $m_e/m_P$. In particle physics terms, the fact that the
depolarization time scale is long is a manifestation of the heavy
quark spin symmetry. In the $m \to \infty$ limit the spin of the heavy
quark is conserved. Thus, it can be changed only on time scales that
are associated with energy scales that are suppressed by at least
$1/m$.

In fact, the loss of polarization is done in several stages. Thus, a
more refined knowledge about the lifetime can be obtained. For the
purpose of illustration we consider a world where the top quark has a
long lifetime. We neglect hadronization into baryons, and consider the
two lightest top-mesons, $T^*$ and $T$ (the analog of $B^*$ and
$B$). These two states form a doublet under the heavy quark spin
symmetry \cite{Manohar:2000dt}. We further consider a very clean
environment where we know the initial top polarization. Then, the
angular distribution of the decay products can be used to measure the
top polarization. Depolarization effects caused by the fact that the
top hadronizes, make the final polarization smaller than that of a
free quark.

There are several time scales associated with hadronization and
depolarization \cite{Falk:1993rf}:
\begin{enumerate}
\item $t_1^{-1} \sim \lqcd$: This is the time scale where hadronization
occurs. That is, after that time the top quark is hadronized into a
heavy hadron, which can be a superposition of $T$ and $T^*$, and
possibly many light hadrons. (There is a small probability to
hadronized into a top baryon, which we neglect for now.) Since the mass
difference between $T$ and $T^*$ is much smaller than $\lqcd$, the
meson containing the top quark is not in a mass eigenstate but rather a
coherent superposition of $T$ and $T^*$.
\item $t_2^{-1} \sim \Delta m$: The next relevant time scale is that
associated with the splitting between the two hadrons
\beq
\Delta m \equiv m(T^*)-m(T).
\eeq
At this time the system starts to ``feel'' the mass difference between
the two hadrons. The system oscillates between the two mass
eigenstates, which practically means loss of coherence. $t_2$ is
the time scale that controls the first depolarization
stage.\footnote{It is often said that ``the top keeps its spin since
it decays before it hadronizes.'' While this statement is correct, it
is misleading. The relevant time scale for depolarization is much
longer than the hadronization time scale.} As we show below, at times
much larger than $t_2$, half of the initial polarization is lost.
\item  $t_3^{-1} \sim \Gamma_{\gamma}$: The last relevant 
time scale is the one that controls the $T^* \to T$ transition
\beq
\Gamma_\gamma \equiv \Gamma(T^*\to T \gamma).
\eeq
Since the $T$ is a scalar, once the $T^*$ decay into a $T$, all the
initial polarization information is lost.
\end{enumerate}
By measuring the amount of depolarization we can get a rough idea
about the width of the top quark, $\Gamma$. If no depolarization
occurs, we know that $\Gamma \gg
\Delta m$. If half of the polarization get lost, it implies that
$\Gamma_{\gamma}\ll\Gamma \ll \Delta m$. All the initial polarization
is lost when $\Gamma \ll \Gamma_\gamma$.

%%%%%%%%%%%%%%%%%%%%%%%%%%%%%%%%%%%%%%%%%%%%%%
\section{Formalism}

We develop the formalism by considering a simple toy model. We
comment about more realistic scenarios later, but a full study of a
realistic model is left for a future work.

Our toy model consists of a heavy ``top'' quark, $t$, a massless
``bottom'' quark, $b$, and a massless scalar, $\phi$.  That is, the
$t$ and the $b$ are spin half fermions that transform as 3 under
SU(3)$_C$ while $\phi$ is a scalar and does not carry color. The
interaction term is chiral
\beq
y_{tb}\,\overline t \,{1-\gamma_5 \over 2}\, b\, \phi.
\eeq
We assume that the top is produced fully polarized and that we know
its spin direction, which we denote as the $z$ axis. We further take
$m_t$ to be known and to be of order a few hundred $\GeV$. In this
simple model we can measure the final top polarization by the angular
dependence of the out going $b$ quark
\beq
{d \Gamma \over d \cos\theta} = {m_t \,y_{tb}^2 \over 64\pi^2}
 \,\left(1-2\vev{s_Z}\cos\theta\right).
\eeq
Here
$\theta$ is the standard azimuthal angle and the normalization is such
that a polarized top has $\vev{s_Z}=1/2$.

We emphasize that the angular distribution of the decay products
depends on the spin of the top quark even after hadronization. It is a
very good approximation to neglect spectator effects in the
decay. Thus, the spin of the hadron is irrelevant in the decay, it is
only the spin of the heavy top that counts.

Once $m_t$ is known, we can use heavy quark symmetry to calculate
$\Delta m$ and $\Gamma_{\gamma}$ (the details of the calculations are
given in the next section). Thus, we assume that the following
quantities are known:
\beq
m_t, \qquad \Delta m\equiv m(T^*)-m(T), \qquad
\Gamma_{\gamma} \equiv \Gamma (T^* \to T \gamma),
\eeq
such that $m_t \gg \Delta m \gg \Gamma_\gamma$. In general,
$\Gamma_{\gamma}$ carries a flavor index, as it depends on the light
quark. Here we further simplify by assuming that the $t$ quark has
only one way to hadronized, say into a $T_d$ meson, and thus we
omitted the flavor index. To a very good approximation both $\Delta m$
and $\Gamma$, the weak decay rate of the top, are independent of the
light degrees of freedom. Therefore, $\Gamma(T)=\Gamma$ and 
width different between the two mesons is
\beq
\Delta \Gamma \equiv \Gamma(T^*)-\Gamma(T)=\Gamma_\gamma.
\eeq

It is useful to define two bases that describe the state of the heavy
meson. The mass basis is spanned by the $T^*$ and $T$ mesons. In this
basis the total spin and the total spin in the $z$
direction are known. The spin basis is the one that is labeled by $s_Z$
of both the top and the spectator $d$ quark.  We denote its eigenvectors by
$\ket{s_t,s_d}$ with $s_t,s_d=+,-$. The relation between the two bases
is
\beqa
T^*(1,1)=\ket{++}, \qquad&&
T^*(1,0)={\ket{-+}+\ket{+-} \over \sqrt{2}}, \nonumber \\ 
T^*(1,-1)=\ket{--}, \qquad&&
T(0,0)={\ket{-+}-\ket{+-} \over \sqrt{2}}.
\eeqa

Next we move to calculate $\vev{s_Z}(t)$, the top polarization as a
function of time. We set $t=0$ as the time the top is produced and
hadronizes. That is, we neglect the stage of hadronization which is
very fast. We assume that the top is produced with a spin in the $z$
direction. We further assume that the light quark in the meson is
unpolarized, that is, it has equal probability to have spin up or
down. See \cite{Falk:1992wt} for discussion on that point. 
Thus, the meson state $T(t)$ at time $t=0$ is an equal incoherent sum
of the following two states
\beq
\ket{++}, \qquad\ket{+-}.
\eeq
In term of the mass eigenstates we have
\beq
\ket{++}=T^*(1,1), \qquad
\ket{+-}= {T^*(1,0) +T(0,0) \over \sqrt{2}} .
\eeq

We assume that $\Delta m$ and $\Delta \Gamma$ are known. Furthermore,
we always have $\Delta m \gg \Delta \Gamma$. Thus,
we can get the time dependences of the two states. Our interested lies
in the 
%\beqa \label{T-t-dep}
%\ket{++}(t)&=&\exp\left(-{(\Gamma+\Delta\Gamma) \,t\over 2}\right)
%T^*(1,1), \nonumber \\
%\ket{+-}(t)&=&\exp\left(-{(\Gamma+\Delta\Gamma) \,t\over 2}\right)
%\left(
%{T^*(1,0)\over \sqrt{2}}+ 
%\exp\left(i\Delta m\,t +{\Delta\Gamma\,t\over 2}\right)
%\,{T(0,0)\over \sqrt{2}}\right)  ,
%\eeqa
%where we omit an irrelevant overall time dependent phase. Using
%Eq.~(\ref{T-t-dep}) we can calculate the expectation value of the 
top polarization as a function of time, $\vev{s_Z}(t)$. We find
\beq \label{T-spin-dep}
{\vev{s_Z}(t) \over \vev{s_Z}(t=0)} = {1 \over 2} \times\left[
\cos(\Delta m \,t) + e^{-\Delta\Gamma t} \right],
\eeq
where we neglect $\Delta \Gamma$ compared to $\Delta m$.

A few points are in order regarding Eq. (\ref{T-spin-dep}):
\begin{enumerate}
\item
At very short times, $t\ll 1/\Delta m$, the polarization is unchanged
from its initial value.
\item
At later times, $1/\Delta\Gamma \gg t \gg 1/\Delta m$ the oscillatory
term is averaged to zero and we see that the polarization reduced to
half its initial value.  This result can be understood from the meson
picture. At $t=0$ the $T^*(1,0)$ and $T(0,0)$ are in a coherent
state. At later times, $t\gg 1/\Delta m$, the state is effectively a
decoherent sum of these two states, each with an average zero top
polarization. The $T^*(1,1)$ state, however, stays polarized, and thus
half of the original polarization is maintained.
\item
At very long times, $t \gg 1/\Delta\Gamma$, when the $T^*\to T\gamma$
decay takes place, there is complete depolarization.
\end{enumerate}

Since in practice the time evolution of the meson cannot be traced, we
have to integrate Eq. (\ref{T-spin-dep}) over time to get the average
polarization. We define $\vev{s_Z}^{free}$ to be the average top
polarization assuming no hadronization, and it is therefore not depend
on time. (We factor out the trivial exponential decay.) We parametrize
the amount of integrated remnant top polarization by
\beq
r\equiv {\int dt \exp(-\Gamma\,t)\,\vev{s_Z}(t) \over 
\int dt  \exp(-\Gamma\,t)\,\vev{s_Z}^{free}}\,,
\eeq
such that $r=1$ indicates that the initial polarization is maintained
while $r=0$ refers to a case that the top decayed after it was
completely depolarized. We find
\beq \label{main-res}
r= {1 \over 2}\left({1 \over 1+x^2}+{1\over 1+y}\right),
\eeq
where we defined
\beq
x\equiv {\Delta m \over \Gamma}, \qquad 
y\equiv {\Delta\Gamma \over 2\Gamma}.
\eeq

\begin{figure}
\begin{center}
\psfrag{r}{$r$}
\psfrag{Gamma}{$\Gamma^{-1} \,[{\rm MeV}^{-1}]$}

\includegraphics[width=9.5cm]{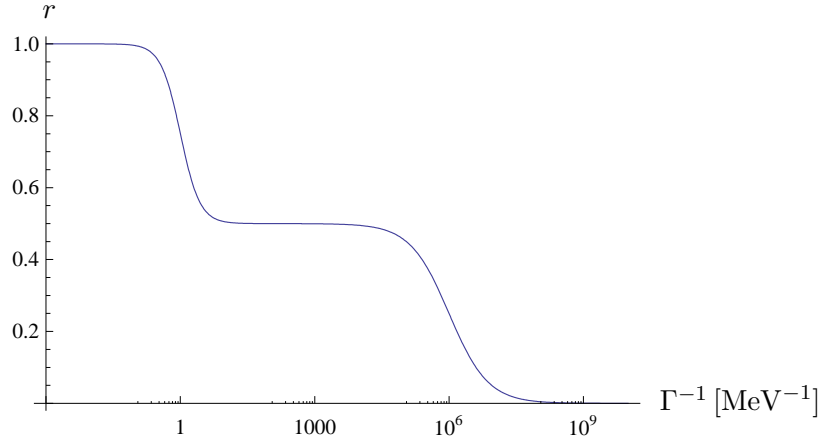}
\end{center}
\centerline{\parbox{14cm}{\caption{\label{fig-simple}
$r$ as a function of $\Gamma^{-1}$ (in MeV$^{-1}$). We use
Eq. (\ref{main-res}) setting $\Delta m=1\,\MeV$ and
$\Delta\Gamma=\Gamma_{\gamma}=1\,\eV$.}}}
\end{figure}

Eq.~(\ref{main-res}) is the main result of this section. It
demonstrates how we can get information about the width of the top,
$\Gamma$. In principle, by measuring $r$ and using $\Delta m$ and
$\Delta\Gamma$ as inputs, we get $\Gamma$ precisely using
Eq.~(\ref{main-res}). This is illustrated in Fig.~\ref{fig-simple}. In
practice, however, the $\Gamma$ dependence of $r$ is strong only near
$x \sim 1$ and $y \sim 1$. Far away from these regions $\Gamma$ cannot
be practically probed. This can be seen in Fig.~\ref{fig-simple} where
far from $x\sim 1$ and $y \sim 1$, $r$ is very flat.

%Integrate[Cos [a x] Exp[-g x],{x,0,Infinity}]
%Integrate[Exp [-a x] Exp[-g x],{x,0,Infinity}] 
%eps = 10^-6; LogLinearPlot[(1/2) ((1/(1 + x^2)) + (1/(1 + x eps))), {x,10^-3, 10 /eps}]
%eps1 = 10^-6; eps2=2*10^-6;eps3=3*10^{-6};LogLinearPlot[(1/2) (0.2+0.9((1/(1 + x^2)) + 0.4(1/(1 + x eps1))+ 0.4(1/(1 + x eps2))+ 0.1(1/(1 + x eps3))), {x,10^-3, 10 /eps}]

%%%%%%%%%%%%%%%%%%%%%%%%%%%%%%%%%%%%%%%%%%%%%%
\section{A more realistic scenario}

In order to make the lifetime probe realistic, the following points
have to be addressed:
\begin{enumerate}
\item
How well can we calculate the initial polarization?
\item
How well can the polarization at the time of decay be measured?
\item
The hadronization can be into a baryon or a meson, and each of them
has several different possibilities for the light degrees of
freedom. How well do we know the flavor ratios after hadronization?
\item
Given the mass, spin, and SU(3)$_C$ representation of the new heavy
particle, how well can we calculate $\Delta m$ and $\Gamma_{\gamma}$?
\end{enumerate}

Regarding the first point. The initial polarization can be calculated
within any specific model. That is, if we have a model we put to test,
in principle, we know the initial polarization. Clearly, the amount of
theoretical uncertainty depends on how well the model parameters are
known. Our hope is that by the time the ideas we propose can be used,
the initial polarization can be determined to high enough precision.

Moving to the second point. Discussion of ways to measure heavy
particles spin and polarization have been studied before, see, for
example, Refs. \cite{Wang:2006hk,Csaki:2007xm,Parke:1996pr}, and for a
recent review see \cite{Wang:2008sw}. The point is that these ideas
can be carried out for the particles we are interested in.

Polarization measurements have been suggested (and used) to determine
the spin of particles. To measure that, those methods do not have to
be very sensitive to the accuracy of the measurement.  To utilize them
for our purpose, however, it is essential to have a good understanding
of both experimental and theoretical errors. We have to address
various questions, for example: Is the angular dependence being washed
out by massive decay products? What the chirality of the decay
vertex is \cite{Wang:2006hk}?

From now on we confine the discussion to the case of a spin half color
triplet particle. We start by determining the flavor ratio after
hadronization. Far from threshold it is reasonable to assume that the
hadronization is independent of the mass of the heavy quark.  Thus, we
can use $b$ data in order to predict the hadronization for a heavy
color triplet spin half. Isospin symmetry tells us that at high energy
the probability to hadronized into $T_u$ and $T_d$ is about the
same. In the $B$ case the two other significant hadronization modes
are into $B_s$ and baryons. Using the data \cite{pdg} it is
straightforward to predict $P(X)$, the probability to hadronized into
the hadron $X$, as
\beq \label{had-ratio}
P(T_u)\approx 40\%,\qquad P(T_d)\approx 40\%, \qquad P(T_s)\approx
10\%, \qquad P(\Lambda_t)\approx 10\%.
\eeq
We use standard notation extended to the top case. That is, $T_q$ is a
meson made of $t$ and and $\bar q$ quark, while $\Lambda_t$ is a
baryon made out of a $t$ and two light quarks.  Note that close to
threshold the situation might be different as phase space effects can
be important. In principle, such effects can be estimated.

Next we discuss the determination of $\Delta m$ and $\Gamma_{\gamma}$.
We start with the baryons. The lowest state, denoted by $\Lambda_t$,
is a spin half. The two light degrees of freedom are in a relative
spin zero configuration, and therefore $\Lambda_t$ is a singlet of the
heavy quark spin symmetry. Since the light degrees of freedom are in
a spin zero configuration the spin of the baryon is the same as the
spin of the heavy quark. Thus, the baryon keeps the initial
polarization and hadronization effects are not important.\footnote{For
the $B$ case there is a subtlety related to the $\Sigma_b$ that can
actually affect the initial polarization. This issue is discussed in
\cite{Falk:1993rf}, where it is shown that the effects are important
for the case of $m \sim 5\,\GeV$ but can be safely neglected for heavy
quarks with $m \gsim 100\,\GeV$.}

The situation with the meson doublet, $T$ and $T^*$, is more
complicated. In the following we estimate $\Delta m$ and
$\Delta\Gamma$. The idea is to use $D$ and $B$ data and Heavy Quark
Effective Theory (HQET) to predict these quantities for a heavy top.

We start with the calculation of $\Delta m$. It is determined by the
HQET $\lambda_2$ parameter \cite{Falk:1992wt}
\beq
\Delta m = {2 \lambda_2 \over m}.
\eeq
$B$ meson data implies $\lambda_2(\mu=m_b) \approx 0.12\,\GeV$. Using
leading log running \cite{Falk:1990pz}, we get
\beq \label{caldm}
\Delta m = \Delta m_B \left({m_B\over m_T}\right)
\left(\alpha_s(m_T)\over \alpha_s(m_B)\right)^{3/(11-2n_f/3)}.
\eeq
Once the new particle mass is determined, we can therefore calculate
$\Delta m$. For example, using $m_t =170\;\GeV$, $n_f=5$ and
$\alpha_s(m_t)/\alpha_s(m_B)\approx 3.5$ we get
\beq
\Delta m \approx 1\;\MeV.
\eeq
While this is only a rough estimate it serves two purposes. First, we
learn that $\Delta m$ is in the range that is of interest to
us. Second, we see that in principle we can get quite an accurate
determination of $\Delta m$. If needed, higher order corrections can
be included.

Next we move to the calculation of $\Gamma_{\gamma}$. We can use heavy
quark symmetry and $D^*$ decay data to get $\Gamma_{\gamma}$ for much
heavier mesons. Following \cite{Amundson:1992yp} and
\cite{Stewart:1998ke} the decay rate can be
parameterized as
\beq \label{ratefor}
\Gamma_{\gamma}^a={\alpha \over 3} |\mu^a|^2 |k_\gamma|^3,
\eeq
where $a=u,d,s$ is the light quark index, $\alpha$ is the fine
structure constant, $k_\gamma$ is the photon momentum, and $\mu^a$ is
a coupling constant of dimension $-1$. We have basically two unknowns
in Eq.~(\ref{ratefor}), $|k_\gamma|$ and $|\mu^a|$. For a very heavy
quark the photon momentum is given by $|k_\gamma|=\Delta m_T$, a
quantity we already discussed, see Eq. (\ref{caldm}). The calculation
of $|\mu^a|$ is more complicated. Both the light and heavy quarks
contribute to $\mu^a$, but their contributions scale like $1/m_q$. For
the $D$ case, where $1/m_c$ is not very small, the charm contribution
is important. In our case, however, since the top is very heavy, we
can neglect its contribution to $\mu^a$. We only need the
contributions of the light quarks in order to calculate $\mu^a$.

To calculate $\mu^a$ we use the approximate flavor SU(3) symmetry. In
the SU(3) symmetry limit $\mu^a$ is proportional to one reduced matrix
element \cite{Amundson:1992yp}
\beq
\mu^a = q_a\beta,
\eeq
where $q_a$ is the electric charge of the light quark, and $\beta$ is
the reduced matrix element. When SU(3) breaking effects are included,
the simple ratio of $2:1:1$ is not maintained. To get a rough estimate
we use here the simple quark model prediction \cite{Amundson:1992yp}
\beqa
\mu_u&=&{2 \over 3}\beta -{g^2 \over 4\pi} 
\left({m_K \over f_K^2 }-{m_\pi \over f_\pi^2}\right) , \nonumber \\
\mu_d&=&-{1 \over 3}\beta -{g^2 \over 4\pi} 
{m_\pi \over f_\pi^2},  \nonumber \\
\mu_s&=&-{1 \over 3}\beta -{g^2 \over 4\pi} 
{m_K \over f_K^2} ,
\eeqa
where $g$ is the effective $T^*\,T\,\pi$ coupling. While at present
our knowledge of the values of $\beta$ and $g$ is limited, if needed
in the future much more precise values can be obtained using the
lattice or updating the analysis of \cite{Stewart:1998ke}. For us it
is enough to use rough values for these parameters. We use the
following representative values
\beq
\beta \sim 3\;\GeV^{-1}, \qquad
g \sim 0.5.
\eeq
(The above values are different from those found in
\cite{Stewart:1998ke}. The reason for it is that the data changed. The
above values are roughly the best fit values using current data
\cite{priv-iain}.)  For $m_t = 170\;\GeV$ and $\Delta m = 1\;\MeV$ we obtain
\beq
\Gamma_{\gamma}^u\approx 1.0\times 10^{-2} \,\eV, \qquad
\Gamma_{\gamma}^d\approx 2.5\times 10^{-3} \,\eV, \qquad
\Gamma_{\gamma}^s\approx 2.5\times 10^{-3} \,\eV.
\eeq
We learn that we can expect $\Delta\Gamma$ of order of
$10^{-2}\,\eV$. This value correspond to lifetimes that are within the
problematic region.

We conclude this section with two comments. First we mention that
$\Delta m$ and $\Delta \Gamma$ scale differently as a function of the
heavy quark.  The leading order scaling is
\beq
\Delta m \propto m_t^{-1}, \qquad
\Delta \Gamma \propto m_t^{-3}.
\eeq
The strong dependence of $y$ on the heavy quark mass make it such that
for very heavy quarks $\Delta\Gamma$ may be very small.

Second we note that $\Delta\Gamma$ is not flavor universal. This is
since the width scales like the square of the light quark electric
charge. The fact that the width is not flavor universal can help us to
get more precise information about the heavy quark lifetime. The point
is that we know the hadronization flavor ratio, and therefore we have
several time scales that control the depolarization. We obtain
\beq \label{main-res-imp}
r=P(\Lambda_t)+{1\over 2}\left[ {1-P(\Lambda_t) \over 1+x^2}+
{P(T_u)\over 1+y_u}+{P(T_d)\over 1+y_d}+{P(T_s)\over 1+y_s}\right].
\eeq
where $y_a\equiv (\Gamma^a_\gamma/2\Gamma)$ and $P(X)$, the hadronization
probability into hadron $X$, are assumed to be known, see
Eq.~(\ref{had-ratio}). We also used $P(\Lambda_t)+
P(T_u)+P(T_d)+P(T_s)=1$. Eq.~(\ref{main-res-imp}) is an improvement
over its simplified version, Eq.~(\ref{main-res}).  We see that it
involves several time scales and thus a refined way to probe $\Gamma$
if it happened to be of the order of $\Gamma_{\gamma}$.

%%%%%%%%%%%%%%%%%%%%%%%%%%%%%%%%%%%%%%%%%%%%%%
\section{Discussions and conclusions}

We discussed only a top like heavy particle, that is, a color triplet
spin half heavy fermion. Our method can be extended to other
representation. Clearly, it can work only for particles that are
charged under the strong interaction and are not scalars. For such
cases, however, it is harder to calculate $\Delta m $ and $\Delta
\Gamma$ since we do not have similar systems that we can use to
extrapolate like we did with the $b$ and $c$ quarks. Yet, we do not
see a fundamental obstacle to calculate it using models for QCD or on
the lattice.

There are several issues that have to be under control before our
method can be used:
\begin{enumerate}
\item
We must know the spin and the SU(3)$_C$
representation of the new particle. 
\item
We must have a reliable
way to calculate its initial polarization. 
\item
We need a way to
measure the polarization when the heavy particle decays.
\end{enumerate}
We do not discuss these issues in detail here. We mention that we
would like to know all that independently of our motivation.  We do,
however, show that in some cases there are ways around some of the
above requirements.

If we work within a given model, we can calculate the first two items.
That is, once all the model parameters are given, we can check if the
apparent measurement of the lifetime using our method agrees with the
model prediction.

There is, in principle, a model independent way to avoid the need to
know the initial polarization. Consider a situation where we can
experimentally separate events where the top hadronizes into a baryon
or a meson. As we discussed, baryonic events have negligible
depolarization, and therefore their final polarization is the same as
the initial polarization of the mesonic events.  Thus, the baryonic
events can be used to measure the initial polarization.

We did not discuss the possibility of spectator decay $s\to u e^-
\overline \nu$.  This decay introduces a new time scale which could in
principle add an additional probe on the lifetime by changing the
polarization in the case of the spectator is an $s$-quark. We did not
study this decay in detail, and only comment that it can be relevant
for heavy particles where $\Gamma_\gamma$ is very small.

To conclude, we show that hadronization can be used to probe lifetimes
of particles with intermediate width. The basic idea is that the
depolarization time depends on known QCD dynamics. For particles with
weak scale masses, the depolarization time happen to be in the
region that corresponds to intermediate lifetimes. Therefore, a
measurement of the amount of depolarization can be used to determine
the lifetime of such a particle.

%%%%%%%%%%%%%%%%%%%%%%%%%%%%%
\section*{Acknowledgments} 
We thank Zoltan Ligeti, Maxim Perelstein, Michael Peskin and Iain
W.~Stewart for helpful discussions. This research is supported by the
NSF grant PHY-0355005.

%%%%%%%%%%%%%%%%%%%%%%%%%%%%%


\begin{thebibliography}{99}

%\cite{Langacker:2008ip}
\bibitem{Langacker:2008ip}
  P.~Langacker, G.~Paz, L.~T.~Wang and I.~Yavin,
  %``Aspects of Z'-mediated Supersymmetry Breaking,''
  arXiv:0801.3693 [hep-ph].
  %%CITATION = ARXIV:0801.3693;%%

%\cite{ArkaniHamed:2004fb}
\bibitem{ArkaniHamed:2004fb}
  N.~Arkani-Hamed and S.~Dimopoulos,
  %``Supersymmetric unification without low energy supersymmetry and  signatures
  %for fine-tuning at the LHC,''
  JHEP {\bf 0506}, 073 (2005)
  [arXiv:hep-th/0405159].
  %%CITATION = JHEPA,0506,073;%%


%\cite{Agashe:2004ci}
\bibitem{Agashe:2004ci}
  K.~Agashe and G.~Servant,
  %``Warped unification, proton stability and dark matter,''
  Phys.\ Rev.\ Lett.\  {\bf 93}, 231805 (2004)
  [arXiv:hep-ph/0403143].
  %%CITATION = PRLTA,93,231805;%%

%\cite{Agashe:2004bm}
\bibitem{Agashe:2004bm}
  K.~Agashe and G.~Servant,
  %``Baryon number in warped GUTs: Model building and (dark matter related)
  %phenomenology,''
  JCAP {\bf 0502}, 002 (2005)
  [arXiv:hep-ph/0411254].
  %%CITATION = JCAPA,0502,002;%%

%\cite{Falk:1993rf}
\bibitem{Falk:1993rf}
  A.~F.~Falk and M.~E.~Peskin,
  %``Production, decay, and polarization of excited heavy hadrons,''
  Phys.\ Rev.\  D {\bf 49}, 3320 (1994)
  [arXiv:hep-ph/9308241].
  %%CITATION = PHRVA,D49,3320;%%

%\cite{Manohar:2000dt}
\bibitem{Manohar:2000dt}
  For a review see, for example, 
  A.~V.~Manohar and M.~B.~Wise,
  ``Heavy quark physics,''
  Camb.\ Monogr.\ Part.\ Phys.\ Nucl.\ Phys.\ Cosmol.\  {\bf 10}, 1 (2000).
  %%CITATION = CMPCE,10,1;%%


%\cite{Falk:1992wt}
\bibitem{Falk:1992wt}
  A.~F.~Falk and M.~Neubert,
  %``Second order power corrections in the heavy quark effective theory. 1.
  %Formalism and meson form-factors,''
  Phys.\ Rev.\  D {\bf 47}, 2965 (1993)
  [arXiv:hep-ph/9209268].
  %%CITATION = PHRVA,D47,2965;%%


%\cite{Wang:2006hk}
\bibitem{Wang:2006hk}
  L.~T.~Wang and I.~Yavin,
  %``Spin Measurements in Cascade Decays at the LHC,''
  JHEP {\bf 0704}, 032 (2007)
  [arXiv:hep-ph/0605296].
  %%CITATION = JHEPA,0704,032;%%



%\cite{Csaki:2007xm}
\bibitem{Csaki:2007xm}
  C.~Csaki, J.~Heinonen and M.~Perelstein,
  %``Testing Gluino Spin with Three-Body Decays,''
  JHEP {\bf 0710}, 107 (2007)
  [arXiv:0707.0014 [hep-ph]].
  %%CITATION = JHEPA,0710,107;%%

%\cite{Parke:1996pr}
\bibitem{Parke:1996pr}
  S.~J.~Parke and Y.~Shadmi,
  %``Spin correlations in top quark pair production at e+ e- colliders,''
  Phys.\ Lett.\  B {\bf 387}, 199 (1996)
  [arXiv:hep-ph/9606419].
  %%CITATION = PHLTA,B387,199;%%

%\cite{Wang:2008sw}
\bibitem{Wang:2008sw}
  L.~T.~Wang and I.~Yavin,
  %``A Review of Spin Determination at the LHC,''
  arXiv:0802.2726 [hep-ph].
  %%CITATION = ARXIV:0802.2726;%%

\bibitem{pdg}
The Review of Particle Physics,
W.-M.~Yao et al., Journal of Physics, G 33, 1 (2006).


%\cite{Falk:1990pz}
\bibitem{Falk:1990pz}
  A.~F.~Falk, B.~Grinstein and M.~E.~Luke,
  %``Leading mass corrections to the heavy quark effective theory,''
  Nucl.\ Phys.\  B {\bf 357}, 185 (1991).
  %%CITATION = NUPHA,B357,185;%%


%\cite{Amundson:1992yp}
\bibitem{Amundson:1992yp}
  J.~F.~Amundson {\it et al.},
  %``Radiative D* decay using heavy quark and chiral symmetry,''
  Phys.\ Lett.\  B {\bf 296}, 415 (1992)
  [arXiv:hep-ph/9209241].
  %%CITATION = PHLTA,B296,415;%%


%\cite{Stewart:1998ke} 
\bibitem{Stewart:1998ke} 
  I.~W.~Stewart, 
  %``Extraction of the D* D pi coupling from D* decays,'' 
  Nucl.\ Phys.\  B {\bf 529}, 62 (1998) 
  [arXiv:hep-ph/9803227]. 
  %%CITATION = NUPHA,B529,62;%% 
 


\bibitem{priv-iain}
I.~W.~Stewart, private communication.



\end{thebibliography}
\end{document}